\documentclass{aa}

\usepackage[varg]{txfonts}
\usepackage{graphicx}
\usepackage[colorlinks=true,linkcolor=magenta,citecolor=blue,filecolor=cyan,urlcolor=red]{hyperref}

\DeclareUnicodeCharacter{2212}{-}
\makeatletter
\renewcommand*\aa@pageof{, page \thepage{} of \pageref*{LastPage}}
\makeatother

\begin{document} 

   \title{Shock-accelerated electrons during the fast expansion of a coronal mass ejection}
        \titlerunning{Shock-accelerated electrons during the fast expansion of a coronal mass ejection}

   \author{D.~E.~Morosan \inst{1}
           \and
        J. Pomoell \inst{1}
        \and
        A.~Kumari \inst{1}
        \and
        R. Vainio \inst{2}
        \and
        E.~K.~J.~Kilpua \inst{1} 
        }

   \institute{Department of Physics, University of Helsinki, P.O. Box 64, FI-00014 Helsinki, Finland \\
              \email{diana.morosan@helsinki.fi}
        \and
             Department of Physics and Astronomy, University of Turku, 20014, Turku, Finland
             }

   \date{Received ; accepted }

 
  \abstract
    {Some of of the most prominent sources for energetic particles in our Solar System are huge eruptions of magnetised plasma from the Sun called coronal mass ejections (CMEs), which usually drive shocks that accelerate charged particles up to relativistic energies. In particular, energetic electron beams can generate radio bursts through the plasma emission mechanism. The main types of bursts associated with CME shocks are type II and herringbone bursts. However, it is currently unknown where early accelerated electrons that produce metric type II bursts and herringbones propagate and when they escape the solar atmosphere. }
    {Here, we investigate the acceleration location, escape, and propagation directions of electron beams during the early evolution of a strongly expanding CME-driven shock wave associated with herrinbgone bursts.}
    {We used ground-based radio observations from the  Nan{\c c}ay Radioheliograph combined with space-based extreme-ultraviolet and white-light observations from the Solar Dynamics Observatory and and the Solar Terrestrial Relations Observatory. We produced a three-dimensional (3D) representation of the electron acceleration locations which, combined with results from magneto-hydrodynamic (MHD) models of the solar corona, was used to investigate the origin of the herringbone bursts observed. }
    {Multiple herringbone bursts are found close to the CME flank in plane-of-sky images. Some of these herringbone bursts have unusual inverted J shapes and opposite drifting herringbones also show opposite senses of circular polarisation. By using a 3D approach combined with the radio properties of the observed bursts, we find evidence that the first radio emission in the CME eruption most likely originates from electrons that initially propagate in regions of low Alfv\'en speeds and along closed magnetic field lines forming a coronal streamer. The radio emission appears to propagate in the same direction as a coronal wave in three dimensions. }
    {The CME appears to inevitably expand into a coronal streamer where it meets ideal conditions to generate a fast shock which, in turn, can accelerate electrons. However, at low coronal heights, the streamer consists of exclusively closed field lines indicating that the early accelerated electron beams do not escape. This is in contrast with electrons which,\ in later stages, escape the corona so that they are detected by spacecraft. }

   \keywords{Sun: corona -- Sun: radio radiation -- Sun: particle emission -- Sun: coronal mass ejections (CMEs) -- Sun: activity -- Sun: flares -- Sun: radio radiation}

\maketitle


\section{Introduction}

{Large eruptions of magnetised plasma on the Sun, that is coronal
mass ejections (CMEs), are often the drivers of collisionless plasma shocks which, in turn, can accelerate particles to high energies. Shocks and the propagation of the high-energy particles they accelerate are important phenomena as they can influence planetary magnetospheres, ionospheres, and atmospheres, and they may also affect spacecraft and astronaut safety \citep{Krivolutsky2012, vainio09}.}    

{Remote signatures of shock-accelerated electrons on the Sun are often observed as bursts of radiation at radio wavelengths produced by the plasma emission mechanism \citep{kl02}. These bursts of radiation belong to a class of solar radio bursts called type II bursts \citep[e.g.][]{ma96,ne85,kumari2017a}, which are usually associated with expanding shocks in the corona \citep[e.g.][]{zu18, mancuso2019, Morosan2020a}. type II bursts consist of emission lanes in dynamic spectra drifting from high to low frequency. These lanes usually have a 2:1 frequency ratio representing emission at the fundamental and harmonic of the plasma frequency. type II bursts can also show a variety of fine frequency structures that compose these emission lanes \citep[][]{magdalenic20}. The most well-known of these fine structures are called `herringbones', which are bursts that can accompany type IIs \citep{ho83,ca87,ca89}. Herringbones appear as fast drifting bursts in dynamic spectra, drifting to high or low frequencies, or both, stemming from a type II lane or `backbone'; however, they are sometimes observed without an accompanying type II structure \citep[][]{ho83,ma05,mo19a}. Herringbone bursts represent signatures of individual electron beams accelerated by a shock predominantly at the flanks of a CME as it expands through the corona in a quasi-perpendicular direction to the surrounding magnetic field \citep{zl93, ca13, mo19a}. Herringbones are generally rare events; they accompany type IIs only 20\% of the time \citep{ca89} and they occur even less often on their own.}

{The shock drift acceleration mechanism is believed to be the main mechanism accelerating type II and herringbone electrons in the low corona (up to 2--3~R$_\odot$ from the Sun centre, where R${_\odot}$ is the solar radius), producing electron beams with energies up to 80~keV \cite[e.g.][]{ma05, mann2018}. However, it is not known where the herrignbone-electron beams propagate following acceleration and if they escape the low corona. Further out in interplanetary space, recent studies have shown that type II bursts occur on the flanks of CMEs in super-critical regions and a quasi-perpendicular shock geometry \citep[e.g.][]{jebaraj2021, hegedus2021}; also, at these heights, the electron beams propagate along open field lines. It is currently unknown if the radio-emitting electron beams in the low corona are later observed in situ by monitoring spacecraft, along with other particles, some with energies reaching up to a few MeV \citep[][]{sandroos2006}. The origin and propagation of herringbone bursts in the low corona is poorly studied in part due to the lack of spectroscopic imaging observations during herringbone events, with only one such event having been imaged before at radio wavelengths \citep{mo19a}.  }

{In this paper, we present the first radio imaging of herringbone bursts at frequencies >150~MHz, and we have combined these observations with magneto-hydrodynamic (MHD) modelling of the solar corona to determine the origin and propagation of these shock-accelerated electron beams. In Sect.~\ref{sec:analysis}, we give an overview of the observations and data analysis techniques used. In Sect.~\ref{sec:results}, we present the results, which are further discussed in Sect.~\ref{sec:discussion}, where we also present our conclusions.}

\begin{figure*}[ht]
    \centering
    \includegraphics[width=0.9\linewidth]{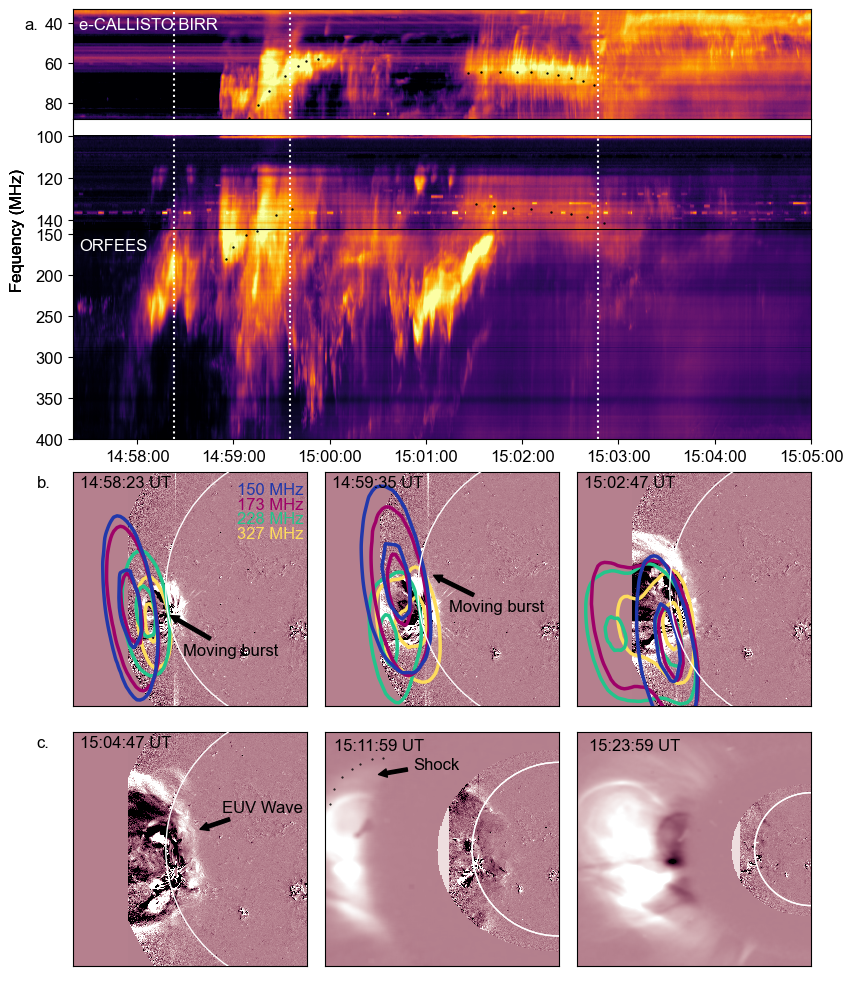}
    \caption{Radio signatures of shock-accelerated electrons associated with a fast-expanding CME. a. Dynamic spectrum from the e-Callisto Birr and ORFEES spectrometers, showing multiple lanes of emission over a wide frequency range consisting of herringbone-radio bursts. b. Early evolution of the CME shown in EUV images from SDO/AIA in the 211~$\AA$ wavelength together with contours of the radio sources at four frequencies: 150, 173, 228, and 327~MHz. c. Evolution of the CME in the outer corona as observed in combined SDO/AIA and white-light images from SOHO/LASCO. A shock is visible as a faint outer boundary in the white-light images.}
    \label{fig:fig1}
\end{figure*}


\section{Observations and data analysis} \label{sec:analysis}

\subsection{Radio emission and associated coronal mass ejection}

\begin{figure*}[ht]
\centering
    \includegraphics[width=0.8\linewidth]{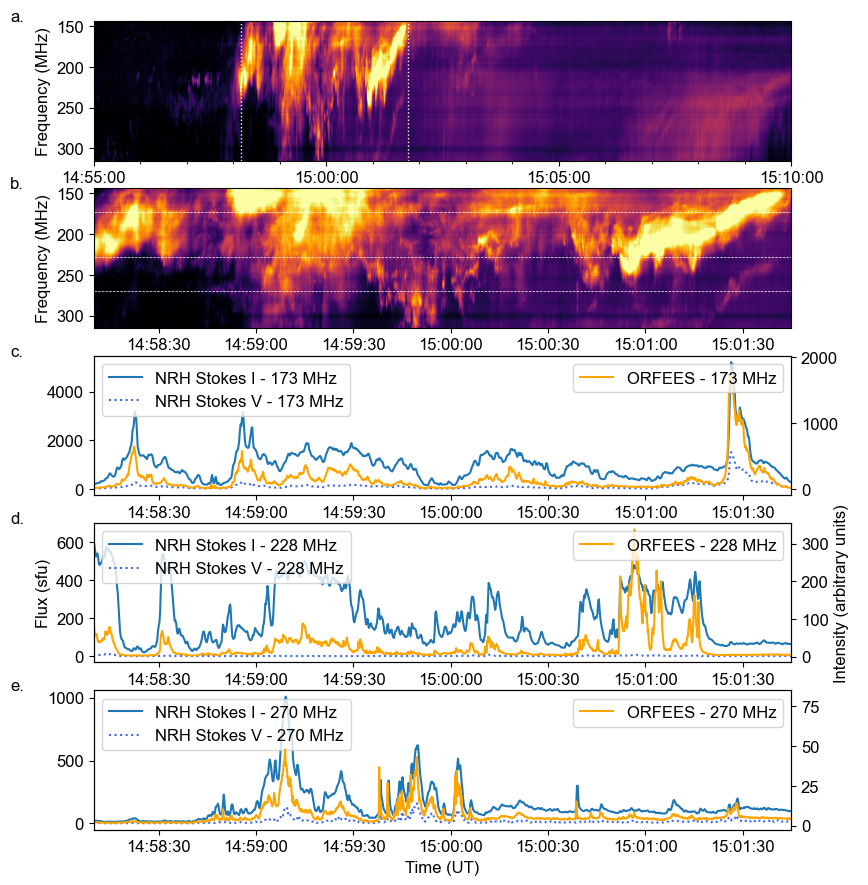}
    \caption{Herringbone dynamic spectrum and their fluxes obtained from NRH images. a. ORFEES dynamic spectrum during the herringbone event. b. Zoom-in of the top panel focussing on groups of herringbones. The last three panels show the flux of the herringbone sources together with the relative intensity extracted from the ORFEES dynamic spectrum at three frequencies: 173 (c), 228 (d), and 270~MHz (e). These plots show that the radio sources imaged by the NRH are indeed the herringbone bursts seen in the ORFEES dynamic spectrum.}
    \label{fig:fig2}
\end{figure*}

{A complex radio event was observed on 25 October 2013 simultaneously with the onset of a fast CME originating close to the solar limb, on the eastern hemisphere. The CME and radio emission accompanied a X-2.1 class solar flare. The CME was observed in remote-sensing observations by multiple spacecraft such as the Solar and Heliospheric Observatory \citep[SOHO;][]{do95, br95} and Solar Dynamics Observatory \citep[SDO;][]{pe12} located in orbits near Earth and the twin Solar Terrestrial Relations Observatory \citep[STEREO;][]{ka08}, orbiting the Sun approximately at Earth's distance (see Fig.~3a). The first observation of the CME in white-light images was at 15:01~UT in STEREO B’s inner coronagraph COR1 \citep{ho08}. A prominent coronal wave was also observed at extreme ultraviolet (EUV) wavelengths by SDO and STEREO, which represents a disturbance of the plasma in the low solar corona ($\sim1.1~$R$_{\odot}$ from the centre) caused by the passage of the CME \citep{long2008, kienreich2009}. Shortly following the onset of the CME, a complex radio event consisting of hundreds of herringbone-radio bursts started at 14:58~UT. These herringbone bursts occurred over a wide frequency range from 10 to 400~MHz and lasted for approximately 20~minutes, indicating the formation of an early low-coronal shock wave \citep{mo19a}. At frequencies >200~MHz, the herringbones are accompanied by a fainter continuum-like emission. }

{The multitude of herringbones observed is shown in Fig.~\ref{fig:fig1} together with the evolution of the CME and herringbone locations. The herringbone bursts are shown in dynamic spectra from the ORFEES spectrograph \citep[][]{Hamini2021}  and the e-Callisto Birr spectrometer \citep{zu12} in Fig.~\ref{fig:fig1}a. Multiple herringbone lanes are observed across the entire frequency range of the dynamic spectrum, 10--400~MHz, some of which show a 2:1 frequency ratio similar to type II bursts, indicating emission at the fundamental and the harmonic of the plasma frequency. This event is similar to the herringbones studied recently by \citet{mo19a} that also occurred without typical type II structures. The event analysed by \citet{mo19a} was associated with one of the largest CMEs observed, with a top speed of $\sim$3000~km/s. The CME related to the herringbones being studied at present is slower, but it can be considered as a fast CME (the top speed is $\sim$1081~km/s as reported by the SOHO Large Angle and Spectrometric Coronagraph (LASCO) CME catalogue, \citealt{yashiro2004}).}

{Radio images were constructed using the Nan{\c c}ay Radioheliograph \citep[NRH;][]{ke97} that observed the Sun until 15:05~UT on the day. Contours of the radio emission sources are shown at four frequencies (150--blue, 173--red, 228--green, and 327~MHz--yellow) in Fig.~\ref{fig:fig1}b, overlaid on running difference images from the Atmospheric Imaging Assembly (AIA) onboard SDO \citep{le12} at a wavelength of 211~\AA. The locations of all herringbones at different frequencies can be seen in Movies 1 and 2 accompanying this paper, which present a full evolution of the eruption.  }

\subsection{Magneto-hydrodynamic modelling of the corona}

{To determine the coronal conditions in the vicinity of herringbone locations, we employed the Magnetohydrodynamics Around a Sphere Thermodynamic (MAST) model \citep{lionello2009}, which can be used to study the large-scale structure and dynamics of the solar corona and inner heliosphere. The MAST model is an MHD model developed by Predictive Sciences Inc.\footnote{http://www.predsci.com/} that uses the magnetic field photospheric magnetograms from the Heliospheric and Magnetic Imager \citep[HMI;][]{sche12} onboard SDO as inner boundary conditions. The model also includes detailed thermodynamic processes with energy equations accounting for thermal conduction parallel to the magnetic field, radiative losses, and coronal heating. This thermodynamic MHD model produces more accurate estimates of plasma density and temperature in the corona and is capable of reproducing global coronal features observed in white light, EUV, and X-ray wavelengths \citep[][]{lionello2009}. In particular, electron densities and magnetic field strengths obtained from the MAST model were used to compute the global Alfv\'en speed in the corona presented in Fig.~5. The low coronal densities have been scaled up by a factor of 2 to obtain more accurate values of their heights \citep{wang17}. }

\begin{figure*}[ht]
\centering
    \includegraphics[width=0.95\linewidth]{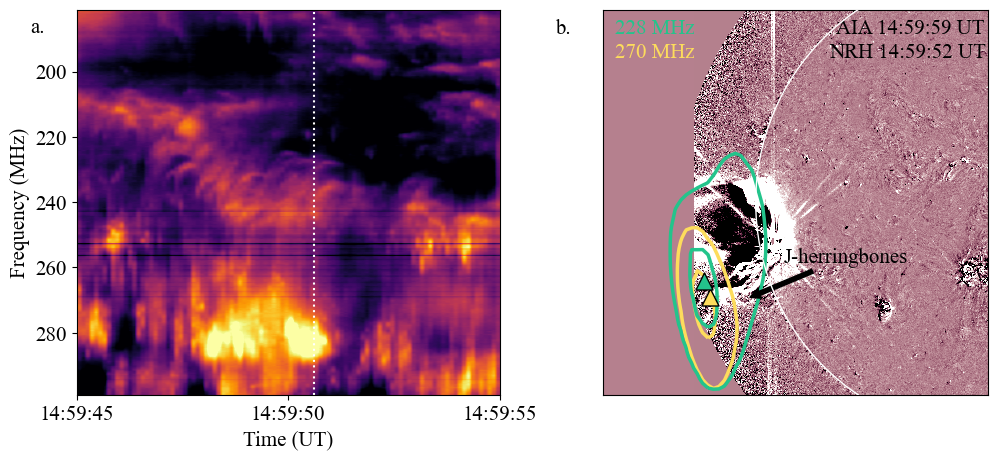}
    \caption{J herringbones and their location relative to the CME. a. Dynamic spectrum showing a group of forward-drifting herringbones with inverted J shapes. b. CME in EUV running difference images from SDO/AIA in the 211~$\AA$ wavelength together with contours of the radio sources at two frequencies: 228 (green) and 270~MHz (yellow). The 228 MHz contours correspond to the tip of a J herringbone. }
    \label{fig:fig3}
\end{figure*}


\section{Results} \label{sec:results}

\subsection{Location and characteristics of the radio emission }

\begin{figure*}[ht]
\centering
    \includegraphics[width=0.85\linewidth]{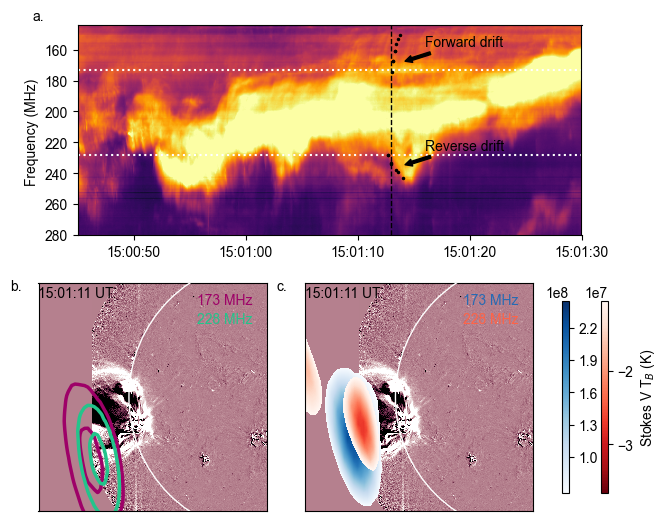}
    \caption{Forward- and reverse-drift herringbones showing opposite senses of circular polarisation. a. ORFEES dynamic spectrum showing herringbones drifting to both high (forward drift) and low frequencies (reverse drift) on either side of a bright backbone. b. AIA 211~\AA~ image of the Sun showing the onset of the CME and overlaid with Stokes I contours of radio sources at 173 and 228~MHz corresponding to the forward- and reverse-drift herringbones, respectively. c. AIA 211~\AA~ image of the Sun showing the onset of the CME and overlaid with Stokes V filled contours of radio sources at 173 and 228~MHz corresponding to the forward- and reverse-drift herringbones, respectively. The colour bars show the brightness temperature of the Stokes V contours. }
    \label{fig:fig4}
\end{figure*}

{The dynamic spectrum in Fig.~\ref{fig:fig1} shows a complex radio event consisting of herringbone emission that does not resemble a typical type II radio burst, which is the usual signature associated with CME shocks. Instead, we observe multiple lanes of emission with herringbones on either side of a backbone (see Fig.~1); however, a bright backbone is not always present. Zoomed-in dynamic spectra of herringbones are shown in Figs.~\ref{fig:fig2}b, \ref{fig:fig3}a and, \ref{fig:fig4}a. The herringbones in the zoomed-in spectra have short durations of $\sim$1~s or less and drift rates of up to $\sim$15~MHz/s, which is similar to values reported in previous studies \citep[][]{ ca13, ca15, mo19a}. Radio imaging with the NRH shows that the herringbones are found at multiple locations relative to the CME (see Fig.~\ref{fig:fig1}b and the movies accompanying this paper) similar to a recent study \citep{mo19a}. }

{To verify the association between the herringbones observed in the ORFEES spectra and the radio sources in NRH images, we computed the flux densities of the moving bursts and compared them to the corresponding ORFEES time series. The flux densities of the moving radio bursts observed by NRH were estimated at 173, 228, and 270~MHz. The flux densities were estimated inside a zoomed-in box covering the full extent and movement of the radio sources, and they include the pixels with levels >20\% of the maximum intensity levels in each box. We chose a threshold of 20\% to include the full extent of the radio source and exclude quiet-Sun or other types of weak emissions. The flux densities were estimated in solar flux units (sfu; where 1~sfu = 10$^{22}$ W m$^{−2}$ Hz$^{-1}$ ) for total intensity (Stokes I) and circularly polarised emission (Stokes V). The NRH flux densities (blue time series in Figs.\ref{fig:fig2}b–c at 173, 228, and 270~MHz) show peaks of bursty emission that correspond to those peaks in the ORFEES normalised intensities at the same frequencies (orange time series in Figs.~\ref{fig:fig2}b–c). Some herringbones also show some circularly polarised emission based on the NRH flux densities and these are discussed in more detail in the following sub-section.}

{One unusual feature found among the herringbones observed is that some of the forward-drifting herringbones show an inverted J structure (see Fig.~\ref{fig:fig3}), indicating that the electrons propagating away from the Sun may encounter closed magnetic loops. The only other radio bursts showing an inverted J shape are J bursts which are a variant of type III bursts propagating along closed magnetic field lines \citep{reid17}. However, J-shaped herringbones have also been reported in a recent study at lower frequencies \citep[][]{magdalenic20}. In our study, such a structure is not seen in the case of the reverse-drift herringbones. The bi-directional electron beams are thus likely to escape in either direction on the same polarity side of a closed magnetic loop, with the forward--drift radio emission likely ending at loop tops (Fig.~\ref{fig:fig3}c).}

\subsection{Circular polarisation of radio emission}

{This herringbone event benefits, for the first time, from simultaneous spectropolarimetric and imaging observations. Here, we present the first images showing the circular polarisation (Stokes V) signal from individual herringbones. The degree of circular polarisation (the ratio of Stokes V to total intensity) of herringbones has been studied on very rare occasions with the main findings being that fundamental emission herringbones were highly polarised (up to 100\%) and all herringbones within a group have the same sense of polarisation \citep{suzuki80}. A zoomed-in dynamic spectrum shows one of the herringbone structures that resembles a single type II-like lane with herringbones on either side of a bright backbone (Fig.~\ref{fig:fig4}a). The backbone contains herringbones with forward (or negative) and reverse (or positive) frequency drift (labelled in Fig.~\ref{fig:fig4}a). The positive frequency drift (the bursts drift from low to high frequency) represents a reverse motion of the electron beam (back towards the Sun) and the negative drift (high to low frequencies) indicates forward motion (away from the Sun). This occurs due to the plasma emission mechanism, where the frequency of emission $f=8980\sqrt{n_e}$ decreases radially from the Sun with decreasing electron density, $n_e$ (here $f$ is in Hz and $n_e$ in cm$^{-3}$). A forward- and a reverse-drift herringbone are shown in Fig.~\ref{fig:fig4}b in total intensity, Stokes  I, contours (30\% and 80\% levels of maximum intensity) and in Fig.~\ref{fig:fig4}c as Stokes V filled brightness temperature ($T_B$) contours. For the first time, we demonstrate that the forward-drift herringbone shows an opposite sense of circular polarisation to the reverse-drift herringbone. Opposite senses of circular polarisation at low radio frequencies have been reported before in the case of other types of radio bursts, indicating propagation of electron beams along open magnetic field lines coming from two different regions of opposite polarities or closed magnetic loops \citep{kai70, Morosan2020a}. Here, the opposite sense of circular polarisation is produced by shock--accelerated bi-directional electron beams (beams travelling towards and away from the Sun) on the same polarity along the magnetic field lines since the forward- and reverse-drift sources occur at roughly the same location.} 

{The majority of herringbones have a low Stokes V signal (see Movie 1 and Fig.~\ref{fig:fig2}) and the degree of circular polarisation is also low (up to $\sim$30\%), in contrast to previously reported values in \citet{suzuki80} for herringbones at frequencies below 160~MHz. The presence of structures with a 2:1 frequency ratio and the low degree of circular polarisation for the herringbones imaged by the NRH indicate that the herringbones observed here at frequencies >150~MHz most likely represent harmonic plasma emission, and the lower frequency herringbones are thus likely to be fundamental plasma emission. }

\subsection{The expansion of the coronal mass ejection and location of herringbones in three dimensions}
 
\begin{figure*}[ht]
\centering
    \includegraphics[width=17cm]{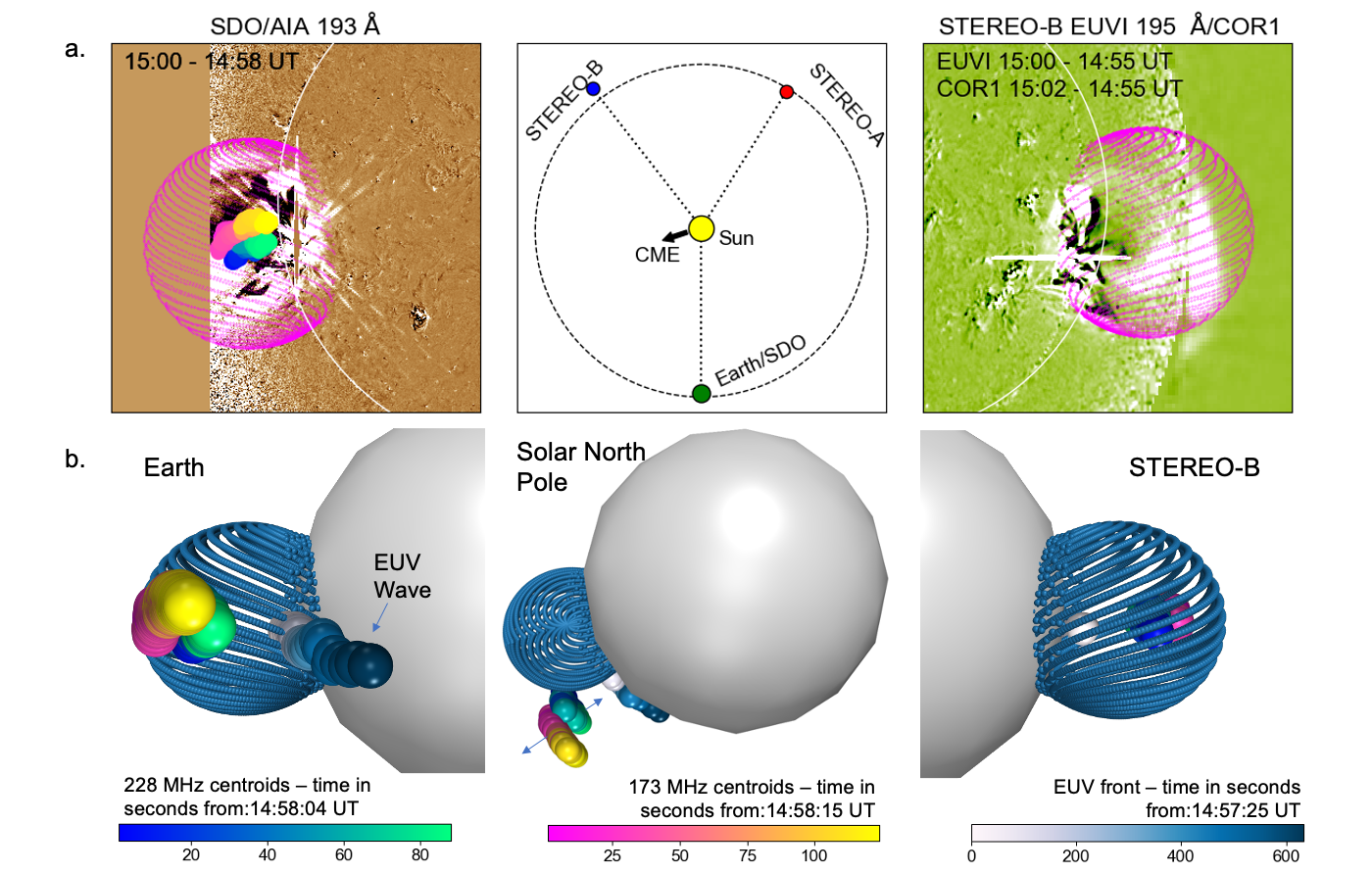} 
    \caption{Three-dimensional reconstruction of the CME eruption and propagation of radio sources. a. Two perspectives of the eruption from the perspective of the SDO spacecraft in orbit near Earth in AIA 193~\AA \ difference images of the Sun (left) and the STEREO-B spacecraft in composite images from EUVI~195~/AA and COR1 images (right). The middle panel shows the locations of these two spacecraft. The STEREO spacecraft were separated from Earth by $148^{\circ}$ (STEREO-A) and $142^{\circ}$ (STEREO-B) in longitude, respectively. The magenta wire mesh represents the reconstruction of the CME bubble and the coloured circles represent the centroids of the moving herringone source. b. 3D reconstruction of the CME eruption together with the centroids of the moving herringbones and propagation of the EUV wave front from three perspectives: Earth (left), solar North Pole (middle), and STEREO-B (right). The colour bars in the bottom represent time in seconds and denote the colouring of the spheres of the 173 and 228~MHz herringbone centroids and one point on the EUV wave front.}
    \label{fig:fig5}
\end{figure*}

{In the plane-of-sky images, the herringbone sources are located at the southern CME flank and also overlap with the CME plasma. However, plane-of-sky projection effects need to be taken into account to accurately determine the position of the herringbones relative to the CME \citep[e.g.][]{chrysaphi20, Morosan2020b}. A single--frequency moving herringbone source is also identified (see the panels of Fig.~\ref{fig:fig1}b and Movies 1 and 2 accompanying this paper). The centroids of this source over time are shown in Fig.~\ref{fig:fig5} and show that the moving source propagates in the direction of the EUV wave towards the solar disc central meridian in the plane of sky (as opposed to movement away from the Sun with the CME). Such a movement indicates that the herringbones propagate at a significant distance outside the plane of sky and it is thus only an apparent movement towards the solar disc in projected 2D images. A 3D approach must be considered to obtain an accurate location of the radio emission relative to the CME, which has been achieved on a few occasions \citep{magdalenic14, mancuso2019, mo19a, chrysaphi20, Morosan2020b, jebaraj2021}. }

{A 3D approach is thus necessary to determine the propagation direction of herringbones relative to the CME and EUV wave. To determine the position of the radio sources in 3D, we de-projected the radio source centroids from the plane-of-sky view. In the low corona, this was done using an electron density model to determine the height of the radio bursts at a specific frequency. This indirect method is primarily required due to the lack of direct observations of the electron densities close to the Sun and limited perspectives of the radio emission (the only perspective available at these frequencies is that of radio telescopes from Earth). In the higher corona and at frequencies below 10~MHz, the availability of multiple spacecraft at various locations around the Sun have allowed for better estimations of the 3D position of radio sources using triangulation methods \citep[e.g.][]{jebaraj2020, hegedus2021, martinez2012, makela2016}. }

{We used the radially-symmetric electron density model of \cite{new61} to estimate the heights of the moving radio sources at a specific frequency. Assuming the radio bursts are emitted at the harmonic of the plasma frequency, the radial distance corresponding to the plasma frequencies of 150 and 173~MHz, respectively, was calculated using a four-fold Newkirk density model. This distance combined with the plane-of-sky coordinates of the centroids was then used to estimate the coordinates of the radio sources in 3D. There is, however, an uncertainty in estimating the radial distances of radio bursts using density models. Firstly, the solar corona is variable and other density models can predict different distances. We assigned an error of $\pm0.3$~$R_\odot$ to the height estimates to reflect a range of possible heights of the radio sources based on two density models, one of the background solar corona \citep{saito77} and a six-fold Newkirk model that reflects the conditions above active regions. }

{The three perspectives from SDO--SOHO and the two STEREO spacecraft allowed us to reconstruct the 3D structure of the CME evolving through time. By employing a geometric model and using the different viewpoints offered by STEREO-B and SDO (see Methods), it is possible to estimate the size and location of the CME plasma. We reconstructed the 3D shape of the CME shock during its early propagation through the solar corona using the graduated cylindrical shell (GCS) model \citep{the06,the09}, which allowed us to fit a parameterised surface to multi-point coronagraph observations. We performed reconstructions using the three viewpoints provided by SDO, LASCO, and the two STEREO spacecraft at 15:24 UT, and, based on these parameters, we reconstructed the CME at earlier times using the EUV data from SDO and EUVI and COR1 data from STEREO-B. The early CME and shock shape was reconstructed at three consecutive times from 15:00 to 15:10~UT. The model was applied to all of the CME emission observed in EUV and white light, including the EUV wave to ensure a more accurate 3D estimation of the extent of the shock. The results of the fitting of the CME at three different times is included in Appendix A. The 3D CME shock (wire frame consisting of magenta dots) is shown from the SDO and STEREO-B perspectives in Fig.~\ref{fig:fig5}a (left and right panels). The middle panel of Fig.~\ref{fig:fig5}a shows the location of the SDO and STEREO-B spacecraft which were placed at ideal locations to provide a side view of the eruption which allows for the reconstruction of the CME in its very early stages. The radio centroids of the moving source labelled in Fig.~\ref{fig:fig1} are shown in the left panel of Fig.~\ref{fig:fig5}a at two frequencies: 173 (pink--yellow dots) and 228~MHz (blue--green dots). The colour of the centroids denote time in seconds and their associated colour bars are shown at the bottom of the figure. }

{The CME and the radio centroids have been reconstructed in 3D in Fig.~\ref{fig:fig5}b from three perspectives: SDO (left), solar North Pole (middle), and STEREO-B (right). One point on the EUV front is also included in this figure as white to blue dots representing time in seconds (labelled EUV wave in the left panel of Fig.~3b). The EUV front points were obtained by tracking the front of the EUV wave in plane-of-sky images close to the moving herringbone source. These points were then de-projected from the plane of sky assuming that the EUV wave travels on a spherical surface at a distance of 1.1~R$_\odot$ from the centre of the Sun \citep{kienreich2009}. The herringbone sources are located outside the reconstructed 3D shock surface (at 15:00~UT) and they expand outwards in the direction of the Earth-facing CME flank. In 3D, the herringbones follow the propagation direction of the EUV wave located at lower heights. The close association in propagation directions of the herringbones and EUV wave strongly suggest that they are both related to the same phenomenon, that is the passage of a CME-driven shock wave or large-amplitude wave from lower to higher heights \citep[e.g.][]{ca13}. The shock wave is theorised to form a dome surrounding the CME \citep{downs12, pomoell2008}, similar to the shape of the reconstructed shock in 3D in Fig.~\ref{fig:fig5}. In the EUV images of the Sun, it is not yet possible to observe the faint outer edge of the CME shock; therefore, the shape of the shock cannot be accurately represented by the fitting methods. The faint outer edge of the shock can usually be identified later in white-light images; however, those images are no longer co-temporal with the herringbones observed (see Fig.~1c). However, our CME reconstruction extends to the outer edge of the EUV wave front and it is likely to include the preceding shock dome. Our reconstruction shows that a shock dome surrounding the CME is likely to be a common origin for both the radio emission and EUV wave. }

\begin{figure*}[ht]
\centering
    \includegraphics[width=0.8\linewidth]{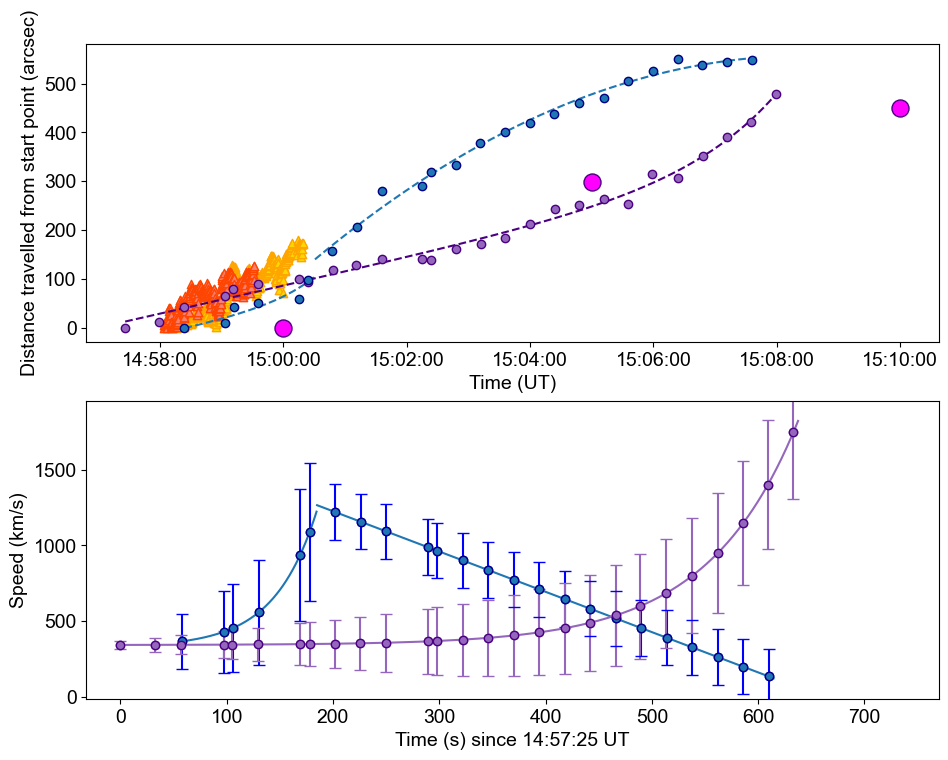}
    \caption{Kinematics of the CME and herringbone bursts. a. Height--time plot showing the propagation of the northern CME flank (blue circles), deprojected EUV wave front (purple circles), herringbone sources at 173 (yellow triangles) and 228~MHz (orange triangles), and deprojected Earth--facing flank (large magenta circles). b. Speed estimates of the northern CME flank and EUV wave front using custom acceleration functions fitted to the height--time points in (a). }
    \label{fig:fig6}
\end{figure*}

{A comparison of the speeds of the herringbone sources, CME and EUV wave (Fig.~\ref{fig:fig5}), reveals that the herringbones are faster than the propagating CME flanks during this early time. Fig.~\ref{fig:fig5}a shows the height--time points of the herringbone centroids at 173 (yellow triangles) and 228~MHz (orange triangles), the EUV wave front (small purple dots), the leading edge of the northern CME flank (small blue dots) and the de-projected 3D CME flank in the same direction as the herringbones sources (large magenta dots). Fig.~\ref{fig:fig5}b shows the fits to the northern CME flank and de-projected EUV wave height--time points. A linear fit to the height--time profile of the herringbone sources (Fig.~\ref{fig:fig5}a) results in a speed estimate of $\sim$1000~km/s at 173~MHz, while a similar fit to the de-projected CME flank gives a lower speed estimate of $\sim$550~km/s. The speed of the de-projected CME flank is based on only three data points due to the low cadence (5~minutes) of STEREO-A data used for the CME reconstruction. While a linear fit is most suitable for the movement of herringbones, the EUV wave and northern CME flank show more complex kinematics that were fitted using  functions containing time--varying acceleration terms. }

{The speeds of the northern CME flank and that of the deprojected EUV wave were estimated by fitting custom functions to their height--time profiles. The northern flank height--time points were estimated by tracking the movement of the outermost edge of the CME in AIA images along an arc at a distance of 1.2~R$_\odot$ from the solar centre. Since the CME is located close to the limb in the SDO perspective, we expect an accurate estimate of the lateral expansion of the CME. The northern flank kinematics show two stages: an early stage of increasing acceleration and a later stage of decreasing acceleration. Thus, the speed was estimated by fitting two custom functions to the height--time data. The EUV wave, throughout the duration considered, shows simpler kinematics with a continuous increasing acceleration. The EUV wave and the beginning of the northern flank kinematics were fitted with a function that contains an exponential acceleration term \citep{ga03}. The height-time function has the following form:
\begin{equation}
h(t) = h_0 + v_0t + a_0 \tau^2 \mathrm{exp}(t/\tau) ,
\end{equation}
where $h_0, v_0$, and $a_0$ are the initial height, velocity, and acceleration, respectively. The function from \cite{ga03} fitted to the CME expansion contains an exponentially varying acceleration of the form $a_0 \mathrm{exp}(t/\tau)$. }

{The later stage of the northern flank expansion was best fitted with a second order polynomial that contains a constant deceleration term, $a_0$: \begin{equation}
h(t) = h_0 + v_0t + \frac{1}{2}a_0 t^2  .
\end{equation}
}       

{The early EUV wave kinematics in Fig.~\ref{fig:fig5} are consistent with an exponential acceleration phase which is typical of the expansion stages of CMEs in the low corona \citep{mo19a,ga03}. However, the kinematics of the northern flank indicates a short period of very fast acceleration, followed by a deceleration phase. The CME shows fast lateral expansion exceeding a speed of 1000~km/s in the early stages of the eruption. The northern flank reaches a top speed of $>1000$~km/s before it begins to decelerate. This occurs during the first few minutes of the eruption, indicating an extremely fast lateral expansion. Comparatively, in the radial direction, the CME reaches a speed of over $1000$~km/s only at a much later stage in the eruption when it reaches the LASCO coronagraphs fields of view (e.g. the last panel of Fig.~1c). This represents evidence that the lateral CME speed (in the direction of the northern flank) is thus faster than the radial CME speed during the early stages of CME expansion. Similarly, in a previous herringbone event, the CME flanks also reach a speed of $\sim$1000~km/s during the early stages of the eruption \citep{mo19a}. The high lateral CME expansion speeds observed in \citet[][]{mo19a} and the current study are likely a significant cause for the formation of a low coronal shock and subsequent radio emission. Moving radio bursts have also been shown to be almost exclusively associated with wide CMEs indicating large expansion speeds in the early stages of the eruption \citep[][]{morosan2021}. The high expansion speeds are also likely to cause super-Alfv\'enic conditions as the CME travels further away from the source active region. The Alfv\'en speed can be investigated either using density models or MHD models of the solar corona. }

\begin{figure*}[ht]
\centering
    \includegraphics[ width=\linewidth ]{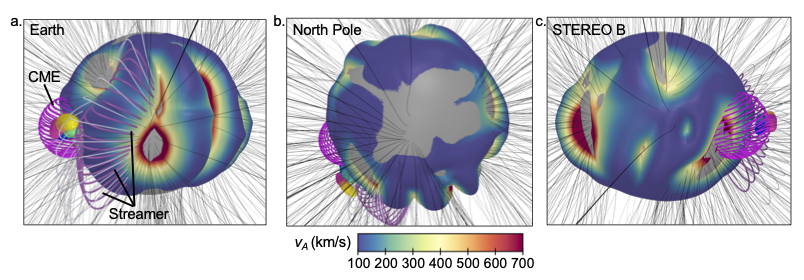}%
    \caption{Modelling of the solar corona during the eruption and herringbone emission. Results of the MAST model showing the 173~MHz harmonic plasma level height, the Alfv\'en speed ($v_A$) at that height, and coronal magnetic field lines from three perspectives: Earth (a), solar North Pole (b), and STEREO B (c). The colour bar at the bottom of the figure represents the scale of the Alfv\'en speed values. Also included in these images are the CME bubble consisting of a magenta wire mesh and the herringbone sources at 173 and 228~MHz using the same colouring as in Fig.~\ref{fig:fig6}. The herringbone sources are roughly located outside the CME and on the side of the closed magnetic field lines that represent a coronal streamer (d).}
    \label{fig:fig7}
\end{figure*}

\begin{figure}[ht]
\centering
    \includegraphics[width=\linewidth, trim = 2cm 4cm 0cm 0cm]{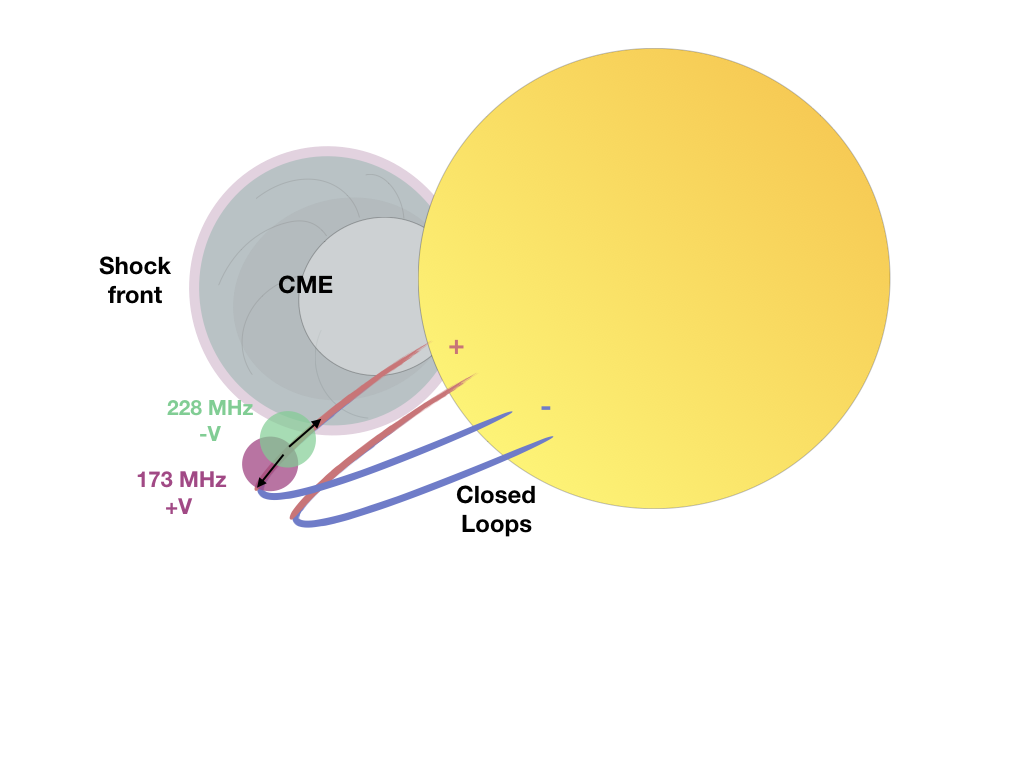}
    \caption{Cartoon showing where the forward- and reverse-drift herringbones are expected to propagate relative to the polarity of the coronal magnetic field.}
    \label{fig:fig8}
\end{figure}

\subsection{Coronal properties in the vicinity of the herringbone sources}

{To determine the properties of the plasma in which the herringbone electrons propagate, we employed the MAST model \citep{lionello2009}. The MAST model outputs used in this study are global electron densities and magnetic field strengths. These properties were then used to compute the global Alfv\'en speed. The results of the model are presented in Fig.~\ref{fig:fig7} which shows electron density iso-contours, magnetic field lines, and the Alfv\'en speed values. The electron density surface in Fig.~\ref{fig:fig7} is $9\times10^7$~cm$^{-3}$, which corresponds to a plasma frequency layer of 175 MHz harmonic emission. The panels of Fig.~\ref{fig:fig7} show the global heights for this plasma density level from three view points: Earth (a), solar North Pole (b), and STEREO B (c). The grey sphere denotes a height of 1.2~R$_\odot$ from the solar centre. The electron density is not uniform in the solar corona, but instead shows ridges of increased density (higher heights of the same density level) at certain locations. Overlaid on these density ridges are values of the Alfv\'en speed where the colour map goes from blue (low Alfv\'en speed) to red (high Alfv\'en speed) and the values are in km/s (a colour bar is included at the bottom of Fig.~\ref{fig:fig7}). Overlaid in the panels of Fig.~\ref{fig:fig7} are open magnetic field lines in black and the closed magnetic field lines (purple and grey). The closed magnetic field lines represent a coronal streamer, which is labelled in Fig.~\ref{fig:fig7}a. The CME shock reconstruction (magenta wire mesh) and the radio centroids from Fig.~\ref{fig:fig5}b are included in these panels, with the same colouring as the colour bars in Fig.~\ref{fig:fig5}b.}

{The MAST model shows that the CME shock and associated radio sources expand into a coronal streamer which, at low heights, is represented by a region of exclusively closed magnetic field lines. The electron densities inside the streamer are also higher compared to the surrounding regions. The streamer region is also a region of low Alfv\'en speeds (as low as $\sim$100~km/s) at the heights in question. The CME expands into the streamer region at a speed of $\sim$550~km/s (based on the deprojected CME flank in Fig.~\ref{fig:fig6}) and, based on the MAST model results, it can become super-Alfv\'enic as soon as it reaches the streamer. Inside the streamer, the CME is capable of driving a shock with a Mach number of $\sim$5, indicating the formation of a strong low coronal shock wave. The northern CME flank was also shown to be very fast with a speed of $\sim$1000~km/s; however, we do not see radio emission associated with the northern flank. This is likely due to the fact that a region of high Alfv\'en speed resides north of the CME source region (see Appendix B).}

{The location of the reconstructed radio sources is ahead of the CME and partly overlapping with the coronal streamer; however, this location is not precise due to uncertainties in de-projecting the radio sources and the model results that are based on a static picture of photospheric conditions at the time. While there is not a perfect positional agreement between the reconstructed radio sources and the streamer location, there is additional evidence that the radio sources expand in an area of closed field lines. Multiple forward drift herringbones in the frequency range 150--270~MHz show an inverted J shape as discussed previously. These herringbones can be seen in the dynamic spectra in Figs.~\ref{fig:fig3}a. Recent studies have shown lower frequency herringbones (10--90~MHz) also with inverted J shapes \citep{magdalenic20}; however, their source locations could not be imaged. }

{There is a limitation in using the MAST model and that is the input magnetogram. The active region producing this eruption is located close to the solar limb in the eastern hemisphere, thus the input magnetogram used for the model run most likely does not show an accurate representation of the surface magnetic field at the time of the eruption due to the obscured view of photospheric features that are located close to the limb. Therefore, the MAST model was only used for the determination of large-scale structures surrounding the active regions; however, we note the exact position of streamers and other features may be slightly offset. Our interpretation can be verified in the future in the case of a similar event where the radio bursts are associated with an active region close to the solar centre, which is likely to yield more accurate results and a better comparison between observations and modelling.}


\section{Discussion and conclusion} \label{sec:discussion}

{The CME studied here shows very fast lateral expansion and it is thus capable of producing a strong low coronal shock.  Once the CME expands into the streamer, based on the MAST results and 3D CME expansion speed, we estimate that the CME is capable of driving a shock with a Mach number of up to $\sim$5. This is the among the highest ever reported Mach number values in the low corona \citep[e.g.][]{mo19a, shen2007}. It has recently been shown that the shock Mach number is the quantity that is most strongly correlated to the peak fluxes of near-relativistic (such as the electron beams studied here) and relativistic electron beams \citep{dresing22}. The high speed of the CME flanks and the common propagation direction of herringbones and the EUV wave supports the idea that a large--scale shock or high--amplitude wave forms ahead of the CME. }

{The CME and likely the related shock wave propagate quasi-perpendicularly to the streamer magnetic field lines and this is the most likely region where herringbones occur. The encounter between the CME shock and streamer in this region allows for the shock--drift acceleration mechanism to produce the electron beams that can generate the observed herringbones. However, due to the presence of the streamer forming closed field lines, the accelerated electrons most likely do not yet escape at low coronal heights. This is also indicated by the inverted J shapes of the herringbones. The inverted J shapes indicate that the electron beams do not propagate along open magnetic field lines, but instead encounter the top of a closed coronal loop.} 

{In the case of J bursts, at the loop top, emission is expected to cease due to increasing electron density gradients when travelling back down the other leg of the loop, which hinders the growth of Langmuir (plasma) waves required for radio emission \citep{reid17}. However, the type III bursts electron beams must travel to large distances before becoming unstable \citep{reid17}, while in this case we are limited by a small acceleration region near the shock and the short travel distances of herringbone bursts (based on their lower frequency bandwidth compared to type IIIs). If a similar mechanism occurs in the case of herringbone bursts, then the propagation through a dense plasma (such as that of a coronal streamer) can result in electron beam instabilities to be achieved quicker that in the case of type III bursts so that the low bandwidth herringbones are produced. The most prominent closed loop, dense plasma system in the propagation direction of the CME is indeed the coronal streamer, adding further supporting evidence that herringbones are likely to be generated in this medium. Unlike some of the lower frequency herringbones (<100 MHz) that are believed to propagate along open field lines \citep{ca13,mo19a}, the high frequency herringbones studied here propagate in closed field-line regions, thus the early accelerated electron beams do not appear to escape into interplanetary space. We present a likely scenario of the generation of herringbone bursts in the cartoon in Fig.~\ref{fig:fig8}, taking into account their sense of circular polarisation and location relative to the CME and streamer. As the CME becomes super-Alfv\'enic close to the streamer, a shock forms which encounters the streamer magnetic field lines quasi-perpendicularly. The shock can then produce bi-directional electron beams that can escape on the same polarity magnetic field lines since the forward- and reverse-drift sources occur at roughly the same location. These electron beams generate herringbones with opposite senses of circular polarisation for forward- and reverse-drift herringbones, and the forward drift herringbones are likely to have inverted J shapes as they encounter magnetic loop tops.}

{Our results open a new interpretation on the origin and propagation of herringbone and type II bursts and have implications on the subsequent acceleration and escape of energetic particles. Electrons generating type II bursts have also recently been reported to originate due to the interaction between streamers and CMEs \citep{fe12, mancuso2019}, with \citet{kouloumvakos21} suggesting that the evolution of a shock inside a streamer plays an important role in the generation of type II emission. Using MHD models of the solar corona, we have shown that the streamer provides ideal conditions (low Alfv\'en speeds, enhanced densities, and closed field regions) for the generation of the J-shaped herringbones, conditions which are also ideal for type II bursts in the same frequency range. CME expansion into coronal streamers is also a common occurrence since streamers are a persistent feature of the solar atmosphere. All these cases \citep{fe12,mancuso2019,kouloumvakos21} where radio signatures are generated inside a streamer indicate that electron beams do no yet escape into interplanetary space. It is likely that shock--accelerated electrons can only escape at larger heights (>2.5~R$_\odot$), where the magnetic field lines are predominantly open, provided that the conditions for acceleration are still favourable. In the present study, there is a clear electron event observed by the STEREO-B High Energy Telescope \citep{rosenvinge08} extending to MeV energies. The onset time of this event is after 15:35~UT, which is clearly after the electron acceleration studied in this paper. These observations of in situ electrons further support the idea that, initially, the herringbone electrons do not escape the solar atmosphere. The shock drift acceleration mechanism alone is not capable of producing high-energy electrons such as the ones observed in situ (the energy gain is only up to a factor of $\sim$4 for strong shocks; e.g. \citealt[][]{takuma2019}). However, multiple shock-crossings can lead to more efficient acceleration to produce MeV electrons. For protons, it has been shown that quasi-perpendicular shocks with curved upstream field lines could lead to the particles being able to interact with the shock many times and that would allow the particles to achieve energies that are much larger than from shock drift acceleration \citep[][]{sandroos2006}. A similar mechanism can also allow electrons to interact with the shock many times to gain higher energies. Future observations that take advantage of the unprecedented fleet of spacecraft in the heliosphere combined with ground-based observations of metric radio bursts could help determine if there is a further link between the electrons generating radio emission at the Sun and those arriving at spacecraft. }




\begin{acknowledgements}{D.E.M acknowledges the Academy of Finland project `RadioCME' (grant number 333859). J.P. acknowledges the Academy of Finland Project 343581. A.K. and D.E.M. acknowledge the University of Helsinki Three Year Grant. E.K.J.K. and A.K. acknowledge the European Research Council (ERC) under the European Union's Horizon 2020 Research and Innovation Programme Project SolMAG 724391. E.K.J.K also acknowledges the Academy of Finland Project 310445. All authors acknowledge the Finnish Centre of Excellence in Research of Sustainable Space (Academy of Finland grant numbers 312390, 312357, 312351  and 336809). This study has received funding from the European Union’s Horizon 2020 research and innovation programme under grant agreement No.\ 101004159 (SERPENTINE). We thank the radio monitoring service at LESIA (Observatoire de Paris) to provide value-added data that have been used for this study. We also thank the Radio Solar Database service at LESIA / USN (Observatoire de Paris) for making the NRH and ORFEES data available. We acknowledge the e-Callisto Birr spectrometer located at the Rosse Observatory and supported by Trinity College Dublin. }\end{acknowledgements}

\bibliographystyle{aa} 
\bibliography{Bib_AA} 

\begin{thebibliography}{60}
\expandafter\ifx\csname natexlab\endcsname\relax\def\natexlab#1{#1}\fi

\bibitem[{{Brueckner} {et~al.}(1995){Brueckner}, {Howard}, {Koomen},
  {Korendyke}, {Michels}, {Moses}, {Socker}, {Dere}, {Lamy}, {Llebaria},
  {Bout}, {Schwenn}, {Simnett}, {Bedford}, \& {Eyles}}]{br95}
{Brueckner}, G.~E., {Howard}, R.~A., {Koomen}, M.~J., {et~al.} 1995, \solphys,
  162, 357

\bibitem[{{Cairns} \& {Robinson}(1987)}]{ca87}
{Cairns}, I.~H. \& {Robinson}, R.~D. 1987, \solphys, 111, 365

\bibitem[{{Cane} \& {White}(1989)}]{ca89}
{Cane}, H.~V. \& {White}, S.~M. 1989, \solphys, 120, 137

\bibitem[{{Carley} {et~al.}(2013){Carley}, {Long}, {Byrne}, {Zucca},
  {Bloomfield}, {McCauley}, \& {Gallagher}}]{ca13}
{Carley}, E.~P., {Long}, D.~M., {Byrne}, J.~P., {et~al.} 2013, Nature Physics,
  9, 811

\bibitem[{{Carley} {et~al.}(2015){Carley}, {Reid}, {Vilmer}, \&
  {Gallagher}}]{ca15}
{Carley}, E.~P., {Reid}, H., {Vilmer}, N., \& {Gallagher}, P.~T. 2015, \aap,
  581, A100

\bibitem[{{Chrysaphi} {et~al.}(2020){Chrysaphi}, {Reid}, \&
  {Kontar}}]{chrysaphi20}
{Chrysaphi}, N., {Reid}, H. A.~S., \& {Kontar}, E.~P. 2020, \apj, 893, 115

\bibitem[{{Domingo} {et~al.}(1995){Domingo}, {Fleck}, \& {Poland}}]{do95}
{Domingo}, V., {Fleck}, B., \& {Poland}, A.~I. 1995, \solphys, 162, 1

\bibitem[{{Downs} {et~al.}(2012){Downs}, {Roussev}, {van der Holst}, {Lugaz},
  \& {Sokolov}}]{downs12}
{Downs}, C., {Roussev}, I.~I., {van der Holst}, B., {Lugaz}, N., \& {Sokolov},
  I.~V. 2012, \apj, 750, 134

\bibitem[{{Dresing} {et~al.}(2022){Dresing}, {Kouloumvakos}, {Vainio}, \&
  {Rouillard}}]{dresing22}
{Dresing}, N., {Kouloumvakos}, A., {Vainio}, R., \& {Rouillard}, A. 2022,
  \apjl, 925, L21

\bibitem[{{Feng} {et~al.}(2012){Feng}, {Chen}, {Kong}, {Li}, {Song}, {Feng}, \&
  {Liu}}]{fe12}
{Feng}, S.~W., {Chen}, Y., {Kong}, X.~L., {et~al.} 2012, \apj, 753, 21

\bibitem[{{Gallagher} {et~al.}(2003){Gallagher}, {Lawrence}, \&
  {Dennis}}]{ga03}
{Gallagher}, P.~T., {Lawrence}, G.~R., \& {Dennis}, B.~R. 2003, \apjl, 588, L53

\bibitem[{{Hamini} {et~al.}(2021){Hamini}, {Auxepaules}, {Bir{\'e}e},
  {Kenfack}, {Kerdraon}, {Klein}, {Lespagnol}, {Masson}, {Coutouly}, {Fabrice},
  \& {Romagnan}}]{Hamini2021}
{Hamini}, A., {Auxepaules}, G., {Bir{\'e}e}, L., {et~al.} 2021, Journal of
  Space Weather and Space Climate, 11, 57

\bibitem[{{Hegedus} {et~al.}(2021){Hegedus}, {Manchester}, \&
  {Kasper}}]{hegedus2021}
{Hegedus}, A.~M., {Manchester}, W.~B., \& {Kasper}, J.~C. 2021, \apj, 922, 203

\bibitem[{{Holman} \& {Pesses}(1983)}]{ho83}
{Holman}, G.~D. \& {Pesses}, M.~E. 1983, \apj, 267, 837

\bibitem[{{Howard} {et~al.}(2008){Howard}, {Moses}, {Vourlidas}, {Newmark},
  {Socker}, {Plunkett}, {Korendyke}, {Cook}, {Hurley}, {Davila}, {Thompson},
  {St Cyr}, {Mentzell}, {Mehalick}, {Lemen}, {Wuelser}, {Duncan}, {Tarbell},
  {Wolfson}, {Moore}, {Harrison}, {Waltham}, {Lang}, {Davis}, {Eyles},
  {Mapson-Menard}, {Simnett}, {Halain}, {Defise}, {Mazy}, {Rochus}, {Mercier},
  {Ravet}, {Delmotte}, {Auchere}, {Delaboudiniere}, {Bothmer}, {Deutsch},
  {Wang}, {Rich}, {Cooper}, {Stephens}, {Maahs}, {Baugh}, {McMullin}, \&
  {Carter}}]{ho08}
{Howard}, R.~A., {Moses}, J.~D., {Vourlidas}, A., {et~al.} 2008, \ssr, 136, 67

\bibitem[{{Jebaraj} {et~al.}(2021){Jebaraj}, {Kouloumvakos}, {Magdalenic},
  {Rouillard}, {Mann}, {Krupar}, \& {Poedts}}]{jebaraj2021}
{Jebaraj}, I.~C., {Kouloumvakos}, A., {Magdalenic}, J., {et~al.} 2021, \aap,
  654, A64

\bibitem[{{Jebaraj} {et~al.}(2020){Jebaraj}, {Magdaleni{\'c}}, {Podladchikova},
  {Scolini}, {Pomoell}, {Veronig}, {Dissauer}, {Krupar}, {Kilpua}, \&
  {Poedts}}]{jebaraj2020}
{Jebaraj}, I.~C., {Magdaleni{\'c}}, J., {Podladchikova}, T., {et~al.} 2020,
  \aap, 639, A56

\bibitem[{{Kai}(1970)}]{kai70}
{Kai}, K. 1970, \solphys, 11, 456

\bibitem[{{Kaiser} {et~al.}(2008){Kaiser}, {Kucera}, {Davila}, {St.~Cyr},
  {Guhathakurta}, \& {Christian}}]{ka08}
{Kaiser}, M.~L., {Kucera}, T.~A., {Davila}, J.~M., {et~al.} 2008, \ssr, 136, 5

\bibitem[{{Katou} \& {Amano}(2019)}]{takuma2019}
{Katou}, T. \& {Amano}, T. 2019, \apj, 874, 119

\bibitem[{{Kerdraon} \& {Delouis}(1997)}]{ke97}
{Kerdraon}, A. \& {Delouis}, J.-M. 1997, in Lecture Notes in Physics, Berlin
  Springer Verlag, Vol. 483, Coronal Physics from Radio and Space Observations,
  ed. G.~{Trottet}, 192

\bibitem[{{Kienreich} {et~al.}(2009){Kienreich}, {Temmer}, \&
  {Veronig}}]{kienreich2009}
{Kienreich}, I.~W., {Temmer}, M., \& {Veronig}, A.~M. 2009, \apjl, 703, L118

\bibitem[{{Klassen} {et~al.}(2002){Klassen}, {Bothmer}, {Mann}, {Reiner},
  {Krucker}, {Vourlidas}, \& {Kunow}}]{kl02}
{Klassen}, A., {Bothmer}, V., {Mann}, G., {et~al.} 2002, \aap, 385, 1078

\bibitem[{{Kouloumvakos} {et~al.}(2021){Kouloumvakos}, {Rouillard}, {Warmuth},
  {Magdalenic}, {Jebaraj}, {Mann}, {Vainio}, \& {Monstein}}]{kouloumvakos21}
{Kouloumvakos}, A., {Rouillard}, A., {Warmuth}, A., {et~al.} 2021, \apj, 913,
  99

\bibitem[{{Krivolutsky} \& {Repnev}(2012)}]{Krivolutsky2012}
{Krivolutsky}, A.~A. \& {Repnev}, A.~I. 2012, Geomagnetism and Aeronomy, 52,
  685

\bibitem[{{Kumari} {et~al.}(2017){Kumari}, {Ramesh}, {Kathiravan}, \&
  {Gopalswamy}}]{kumari2017a}
{Kumari}, A., {Ramesh}, R., {Kathiravan}, C., \& {Gopalswamy}, N. 2017, \apj,
  843, 10

\bibitem[{{Lemen} {et~al.}(2012){Lemen}, {Title}, {Akin}, {Boerner}, {Chou},
  {Drake}, {Duncan}, {Edwards}, {Friedlaender}, {Heyman}, {Hurlburt}, {Katz},
  {Kushner}, {Levay}, {Lindgren}, {Mathur}, {McFeaters}, {Mitchell}, {Rehse},
  {Schrijver}, {Springer}, {Stern}, {Tarbell}, {Wuelser}, {Wolfson}, {Yanari},
  {Bookbinder}, {Cheimets}, {Caldwell}, {Deluca}, {Gates}, {Golub}, {Park},
  {Podgorski}, {Bush}, {Scherrer}, {Gummin}, {Smith}, {Auker}, {Jerram},
  {Pool}, {Soufli}, {Windt}, {Beardsley}, {Clapp}, {Lang}, \& {Waltham}}]{le12}
{Lemen}, J.~R., {Title}, A.~M., {Akin}, D.~J., {et~al.} 2012, \solphys, 275, 17

\bibitem[{{Lionello} {et~al.}(2009){Lionello}, {Linker}, \&
  {Miki{\'c}}}]{lionello2009}
{Lionello}, R., {Linker}, J.~A., \& {Miki{\'c}}, Z. 2009, \apj, 690, 902

\bibitem[{{Long} {et~al.}(2008){Long}, {Gallagher}, {McAteer}, \&
  {Bloomfield}}]{long2008}
{Long}, D.~M., {Gallagher}, P.~T., {McAteer}, R.~T.~J., \& {Bloomfield}, D.~S.
  2008, \apjl, 680, L81

\bibitem[{{Magdaleni{\'c}} {et~al.}(2020){Magdaleni{\'c}}, {Marqu{\'e}},
  {Fallows}, {Mann}, {Vocks}, {Zucca}, {Dabrowski}, {Krankowski}, \&
  {Melnik}}]{magdalenic20}
{Magdaleni{\'c}}, J., {Marqu{\'e}}, C., {Fallows}, R.~A., {et~al.} 2020, \apjl,
  897, L15

\bibitem[{{Magdaleni{\'c}} {et~al.}(2014){Magdaleni{\'c}}, {Marqu{\'e}},
  {Krupar}, {Mierla}, {Zhukov}, {Rodriguez}, {Maksimovi{\'c}}, \&
  {Cecconi}}]{magdalenic14}
{Magdaleni{\'c}}, J., {Marqu{\'e}}, C., {Krupar}, V., {et~al.} 2014, \apj, 791,
  115

\bibitem[{{M{\"a}kel{\"a}} {et~al.}(2016){M{\"a}kel{\"a}}, {Gopalswamy},
  {Reiner}, {Akiyama}, \& {Krupar}}]{makela2016}
{M{\"a}kel{\"a}}, P., {Gopalswamy}, N., {Reiner}, M.~J., {Akiyama}, S., \&
  {Krupar}, V. 2016, \apj, 827, 141

\bibitem[{{Mancuso} {et~al.}(2019){Mancuso}, {Frassati}, {Bemporad}, \&
  {Barghini}}]{mancuso2019}
{Mancuso}, S., {Frassati}, F., {Bemporad}, A., \& {Barghini}, D. 2019, \aap,
  624, L2

\bibitem[{{Mann} \& {Klassen}(2005)}]{ma05}
{Mann}, G. \& {Klassen}, A. 2005, \aap, 441, 319

\bibitem[{{Mann} {et~al.}(1996){Mann}, {Klassen}, {Classen}, {Aurass},
  {Scholz}, {MacDowall}, \& {Stone}}]{ma96}
{Mann}, G., {Klassen}, A., {Classen}, H.~T., {et~al.} 1996, \aaps, 119, 489

\bibitem[{{Mann} {et~al.}(2018){Mann}, {Melnik}, {Rucker}, {Konovalenko}, \&
  {Brazhenko}}]{mann2018}
{Mann}, G., {Melnik}, V.~N., {Rucker}, H.~O., {Konovalenko}, A.~A., \&
  {Brazhenko}, A.~I. 2018, \aap, 609, A41

\bibitem[{{Mart{\'\i}nez Oliveros} {et~al.}(2012){Mart{\'\i}nez Oliveros},
  {Raftery}, {Bain}, {Liu}, {Krupar}, {Bale}, \& {Krucker}}]{martinez2012}
{Mart{\'\i}nez Oliveros}, J.~C., {Raftery}, C.~L., {Bain}, H.~M., {et~al.}
  2012, \apj, 748, 66

\bibitem[{{Morosan} {et~al.}(2019){Morosan}, {Carley}, {Hayes}, {Murray},
  {Zucca}, {Fallows}, {McCauley}, {Kilpua}, {Mann}, {Vocks}, \&
  {Gallagher}}]{mo19a}
{Morosan}, D.~E., {Carley}, E.~P., {Hayes}, L.~A., {et~al.} 2019, Nat. Astron.,
  3, 452

\bibitem[{{Morosan} {et~al.}(2021){Morosan}, {Kumari}, {Kilpua}, \&
  {Hamini}}]{morosan2021}
{Morosan}, D.~E., {Kumari}, A., {Kilpua}, E.~K.~J., \& {Hamini}, A. 2021, \aap,
  647, L12

\bibitem[{{Morosan} {et~al.}(2020{\natexlab{a}}){Morosan}, {Palmerio}, {Lynch},
  \& {Kilpua}}]{Morosan2020a}
{Morosan}, D.~E., {Palmerio}, E., {Lynch}, B.~J., \& {Kilpua}, E.~K.~J.
  2020{\natexlab{a}}, \aap, 633, A141

\bibitem[{{Morosan} {et~al.}(2020{\natexlab{b}}){Morosan}, {Palmerio},
  {Pomoell}, {Vainio}, {Palmroth}, \& {Kilpua}}]{Morosan2020b}
{Morosan}, D.~E., {Palmerio}, E., {Pomoell}, J., {et~al.} 2020{\natexlab{b}},
  \aap, 635, A62

\bibitem[{{Nelson} \& {Melrose}(1985)}]{ne85}
{Nelson}, G.~J. \& {Melrose}, D.~B. 1985, in IN: Solar radiophysics: Studies of
  emission from the sun at metre wavelengths. Cambridge and New York, Cambridge
  University Press., ed. D.~J. {McLean} \& N.~R. {Labrum}, 333--359

\bibitem[{{Newkirk}(1961)}]{new61}
{Newkirk}, G.~J. 1961, \apj, 133, 983

\bibitem[{{Pesnell} {et~al.}(2012){Pesnell}, {Thompson}, \&
  {Chamberlin}}]{pe12}
{Pesnell}, W.~D., {Thompson}, B.~J., \& {Chamberlin}, P.~C. 2012, \solphys,
  275, 3

\bibitem[{{Pomoell} {et~al.}(2008){Pomoell}, {Vainio}, \&
  {Kissmann}}]{pomoell2008}
{Pomoell}, J., {Vainio}, R., \& {Kissmann}, R. 2008, \solphys, 253, 249

\bibitem[{{Reid} \& {Kontar}(2017)}]{reid17}
{Reid}, H. A.~S. \& {Kontar}, E.~P. 2017, \aap, 606, A141

\bibitem[{{Saito} {et~al.}(1977){Saito}, {Poland}, \& {Munro}}]{saito77}
{Saito}, K., {Poland}, A.~I., \& {Munro}, R.~H. 1977, \solphys, 55, 121

\bibitem[{{Sandroos} \& {Vainio}(2006)}]{sandroos2006}
{Sandroos}, A. \& {Vainio}, R. 2006, \aap, 455, 685

\bibitem[{{Scherrer} {et~al.}(2012){Scherrer}, {Schou}, {Bush}, {Kosovichev},
  {Bogart}, {Hoeksema}, {Liu}, {Duvall}, {Zhao}, {Title}, {Schrijver},
  {Tarbell}, \& {Tomczyk}}]{sche12}
{Scherrer}, P.~H., {Schou}, J., {Bush}, R.~I., {et~al.} 2012, \solphys, 275,
  207

\bibitem[{{Shen} {et~al.}(2007){Shen}, {Wang}, {Ye}, {Zhao}, {Gui}, \&
  {Wang}}]{shen2007}
{Shen}, C., {Wang}, Y., {Ye}, P., {et~al.} 2007, \apj, 670, 849

\bibitem[{{Suzuki} {et~al.}(1980){Suzuki}, {Stewart}, \& {Magun}}]{suzuki80}
{Suzuki}, S., {Stewart}, R.~T., \& {Magun}, A. 1980, in Radio Physics of the
  Sun, ed. M.~R. {Kundu} \& T.~E. {Gergely}, Vol.~86, 241--245

\bibitem[{{Thernisien} {et~al.}(2009){Thernisien}, {Vourlidas}, \&
  {Howard}}]{the09}
{Thernisien}, A., {Vourlidas}, A., \& {Howard}, R.~A. 2009, \solphys, 256, 111

\bibitem[{{Thernisien} {et~al.}(2006){Thernisien}, {Howard}, \&
  {Vourlidas}}]{the06}
{Thernisien}, A.~F.~R., {Howard}, R.~A., \& {Vourlidas}, A. 2006, \apj, 652,
  763

\bibitem[{{Vainio} {et~al.}(2009){Vainio}, {Desorgher}, {Heynderickx},
  {Storini}, {Fl{\"u}ckiger}, {Horne}, {Kovaltsov}, {Kudela}, {Laurenza},
  {McKenna-Lawlor}, {Rothkaehl}, \& {Usoskin}}]{vainio09}
{Vainio}, R., {Desorgher}, L., {Heynderickx}, D., {et~al.} 2009, \ssr, 147, 187

\bibitem[{{von Rosenvinge} {et~al.}(2008){von Rosenvinge}, {Reames}, {Baker},
  {Hawk}, {Nolan}, {Ryan}, {Shuman}, {Wortman}, {Mewaldt}, {Cummings}, {Cook},
  {Labrador}, {Leske}, \& {Wiedenbeck}}]{rosenvinge08}
{von Rosenvinge}, T.~T., {Reames}, D.~V., {Baker}, R., {et~al.} 2008, \ssr,
  136, 391

\bibitem[{{Wang} {et~al.}(2017){Wang}, {Reginald}, {Davila}, {St. Cyr}, \&
  {Thompson}}]{wang17}
{Wang}, T., {Reginald}, N.~L., {Davila}, J.~M., {St. Cyr}, O.~C., \&
  {Thompson}, W.~T. 2017, \solphys, 292, 97

\bibitem[{{Yashiro} {et~al.}(2004){Yashiro}, {Gopalswamy}, {Michalek}, {St.
  Cyr}, {Plunkett}, {Rich}, \& {Howard}}]{yashiro2004}
{Yashiro}, S., {Gopalswamy}, N., {Michalek}, G., {et~al.} 2004, Journal of
  Geophysical Research (Space Physics), 109, A07105

\bibitem[{{Zlobec} {et~al.}(1993){Zlobec}, {Messerotti}, {Karlicky}, \&
  {Urbarz}}]{zl93}
{Zlobec}, P., {Messerotti}, M., {Karlicky}, M., \& {Urbarz}, H. 1993, \solphys,
  144, 373

\bibitem[{{Zucca} {et~al.}(2012){Zucca}, {Carley}, {McCauley}, {Gallagher},
  {Monstein}, \& {McAteer}}]{zu12}
{Zucca}, P., {Carley}, E.~P., {McCauley}, J., {et~al.} 2012, \solphys, 280, 591

\bibitem[{{Zucca} {et~al.}(2018){Zucca}, {Morosan}, {Rouillard}, {Fallows},
  {Gallagher}, {Magdalenic}, {Klein}, {Mann}, {Vocks}, {Carley}, {Bisi},
  {Kontar}, {Rothkaehl}, {Dabrowski}, {Krankowski}, {Anderson}, {Asgekar},
  {Bell}, {Bentum}, {Best}, {Blaauw}, {Breitling}, {Broderick}, {Brouw},
  {Br{\"u}ggen}, {Butcher}, {Ciardi}, {de Geus}, {Deller}, {Duscha},
  {Eisl{\"o}ffel}, {Garrett}, {Grie{\ss}meier}, {Gunst}, {Heald}, {Hoeft},
  {H{\"o}randel}, {Iacobelli}, {Juette}, {Karastergiou}, {van Leeuwen},
  {McKay-Bukowski}, {Mulder}, {Munk}, {Nelles}, {Orru}, {Paas}, {Pand ey},
  {Pekal}, {Pizzo}, {Polatidis}, {Reich}, {Rowlinson}, {Schwarz}, {Shulevski},
  {Sluman}, {Smirnov}, {Sobey}, {Soida}, {Thoudam}, {Toribio}, {Vermeulen},
  {van Weeren}, {Wucknitz}, \& {Zarka}}]{zu18}
{Zucca}, P., {Morosan}, D.~E., {Rouillard}, A.~P., {et~al.} 2018, \aap, 615,
  A89

\end{thebibliography}

\begin{appendix} 

\section{Stereoscopic reconstruction of the CME shock shape} \label{app:a}

   \begin{figure*}[]
   \centering
          \includegraphics[width=15cm]{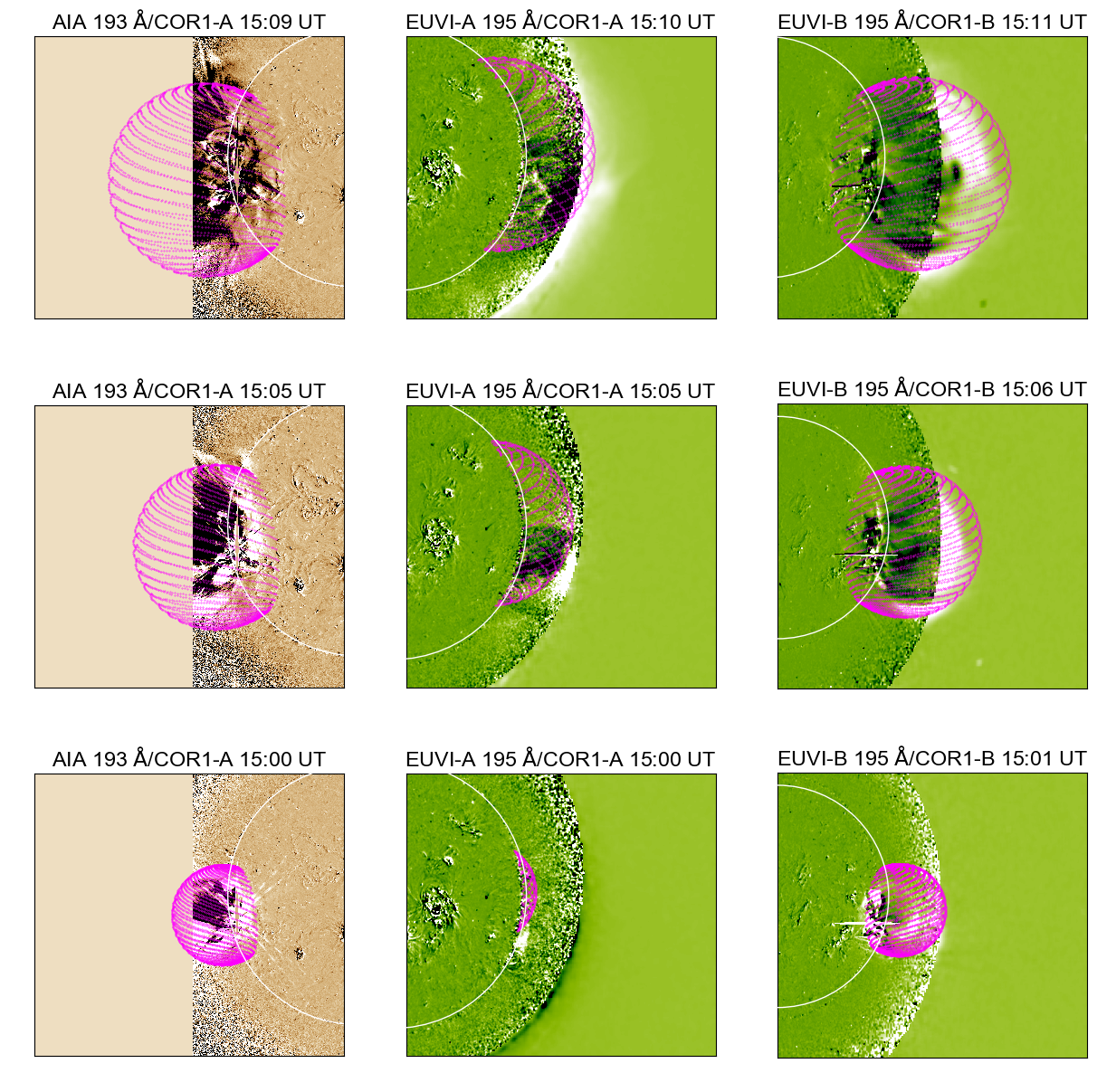}
      \caption{Fitting of the CME with the graduated cylindrical shell (GCS) model at three different times and from three different perspectives: Earth (left), STEREO-A (middle), and STEREO-B (right). The CME fit was done firstly at 15:24~UT when the CME was first visible in the LASCO C2 field of view. Then the fitting parameters were used to estimate the fit at earlier times when the CME was observed only in the Earth at EUV wavelengths and STEREO-A perspectives. The left panels show AIA 193~\AA~ images of the Sun, while the middle and right panels show composite EUVI 195~\AA~ together with COR1 white--light images from the STEREO-A and STEREO-B perspectives, respectively. The magenta wire mesh is the resulting CME fit using the GCS model.}
         \label{figA1}
   \end{figure*}

{We reconstructed the 3D shape of the CME shock during its early propagation through the solar corona using the graduated cylindrical shell (GCS) model \citep{the06,the09}, which allows a parameterised croissant-shaped surface to be fitted to multi-point coronagraph observations. We performed reconstructions using the three viewpoints provided by SDO, LASCO, and the two STEREO spacecraft at 15:24 UT and based on these parameters we reconstructed the CME at earlier times using the EUV data from SDO and EUVI and COR1 data from STEREO-B. The early CME shape was reconstructed at three consecutive times from 15:00 to 15:10~UT. The model was applied to most of the CME emission observed in EUV and white light, including the CME cavity, bright rim, and EUV wave. We chose the GCS model as it better fitted the shape of the entire CME plasma and EUV wave during the early stages of the eruption. The results of the fitting of the CME at three different times is shown in Fig.~\ref{figA1}.}

\section{Magneto-hydrodynamic model of the corona} \label{app:b}

   \begin{figure*}[]
   \centering
          \includegraphics[ width=18cm ]{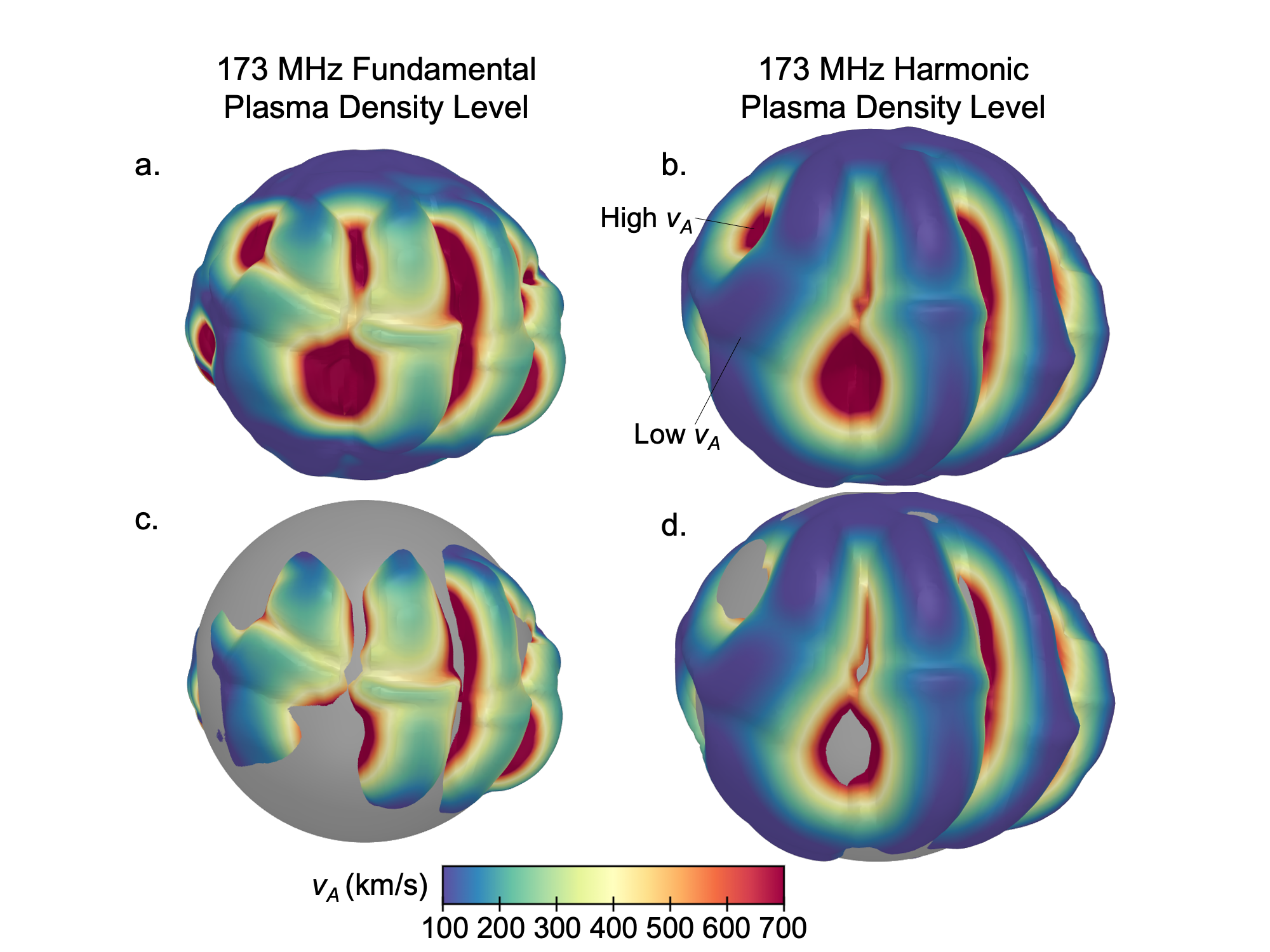} 
      \caption{Electron density and Alfv\'en speeds obtained from the MAST model. Iso-surfaces of the 228 MHz fundamental (a and c) and harmonic plasma level (b and d) with overlaid Alfv\'en speed values at that height obtained from the MAST model. The electron densities corresponding to this surface are $6.4\times10^8$ for fundamental emission and $1.6\times10^8$~cm$^{-3}$ for harmonic emission. The surfaces show the relative height of this emission at various locations on the Sun. The surfaces are overlaid with Alfv\'en speed values at those heights. In the bottom panels, a sphere has been included as a visual guide at a distance of 1.1~R$_\odot$ for fundamental emission (c) and 1.2~R$_\odot$ for harmonic emission (d). The Alfv\'en speed is highly variable across these density surfaces.}
         \label{figB1}
   \end{figure*}

{The MAST model is an MHD model developed by Predictive Sciences Inc.\footnote{http://www.predsci.com/} that uses the magnetic field photospheric magnetograms from HMI onboard SDO as inner boundary conditions. The model also includes detailed thermodynamic processes with energy equations accounting for thermal conduction parallel to the magnetic field, radiative losses, and coronal heating. This thermodynamic MHD model produces more accurate estimates of plasma density and temperature in the corona and is capable of reproducing global coronal features observed in white light, EUV, and X-ray wavelengths \citep[][]{lionello2009}. Electron densities and magnetic field strengths were obtained from the MAST model and were used to compute the global Alfv\'en speed in the corona presented in Fig.~\ref{figB1}. The low coronal densities have been scaled up by a factor of 2 to obtain more accurate values of their heights \citep{wang17}.}

\end{appendix} 

\end{document}